# Millimeter-scale and large-angle self-collimation in a photonic crystal composed of silicon nanorods

Hao Li, [1;2]   Aimin Wu,[1]   Wei Li, [1*]   Xulin Lin, [1]   Chao Qiu, [1;2]   Zhen Sheng, [1]   Xi Wang, [1] Shichang Zou,[1] and Fuwan Gan[1**]

1. State Key Laboratory of Functional Materials for Informatics, Shanghai Institute of Microsystem and Information Technology, Chinese Academy of Sciences, Shanghai 200050, China
2. University of Chinese Academy of Sciences, Beijing 100049, China

*Corresponding author: waylee@mail.sim.ac.cn (Wei Li)
**Corresponding author: fuwan@mail.sim.ac.cn (Fuwan Gan)

**Abstract:** We report the observation of a large-angle self-collimation phenomenon occurring in photonic crystals composed of nanorods. Electromagnetic waves incident onto such photonic crystals from directions covering a wide-range of incident angles become highly localized along a single array of rods, which results in narrow-beam propagation without divergence. A propagation length of 0.4 mm is experimentally observed over the wavelength range of 1540 nm to 1570 nm, even in the large incident angle case, which is a very considerable length scale for on-chip optical interconnection.

## 1. Introduction

Beam spreading arising from a geometrical origin can be manipulated in many ways by using index gradients, delicately designed waveguide arrays [**1**], or nonlinear optical effects [**2**], [**3**]. Photonic crystals (PCs) provide an alternative method for beam steering based on the formation of a photonic bandgap (PBG) and the related dispersion phenomenon, such as negative refraction, superprism, and slow light. Self-collimation (SC), another dispersion-related phenomenon, is also attractive because of its significant impact on channel-less waveguiding [**4**], diffraction inhibition [**5**], subwavelength focusing or imaging [**6**], [**7**], etc.

In recent years, the development of fabrication techniques has enabled PCs to function at optical frequencies, and various self-collimation-based devices have been proposed, such as non-channel waveguides, beam splitters, open cavities [**8**], beam combiners and optic logic [**9**]. Theoretically, many reports in the literature have focused on the conventional and all-angle self-collimation with a flat equi-frequency contour (EFC) across the entire Brillouin zone [**5**], [**10**], [**11**], [**12**], [**13**]. These theoretical studies suggested that an all-angle phenomenon exhibits two unique properties, in contrast to the conventional self-collimation. First, the light waves can be accepted from all incident angles, even up to 90° [12]. Second, the electromagnetic energy can be highly localized without diffraction [5]. For hole-type PCs, the all-angle SC phenomenon will help to confine light waves between air holes; while for pillar-type PCs, the electromagnetic energy can be highly localized along a path as narrow as a single nanorod array, as described in this paper.

Experimentally, the self-collimation phenomenon has been demonstrated in different types of structures, including two-dimensional (2D) hole-type and pillar-type PCs [4], [14], [15], quasi-zero-average-index structures [16], or even three-dimensional PCs[?]. Very recently, a wide-angle SC phenomenon was reported in PCs of square lattices composed of elliptical air holes [17]. Technically speaking, in contrast to the hole-type PCs, the pillar-type structure has significant potential for use with active components because of the possibility of electrical contacting and the heat dissipation capability. Moreover, the rods are more suitable for sensing applications because light interacts with the medium (to be sensed) that surrounds the rods more strongly, and in optofluidics, the rod structure may allow for better fluid penetration than the structure consisting of air holes in a dielectric slab. However, no experimental work has currently been performed on large-angle self-collimation in PCs composed of silicon nanorods, which motivates the current research.

In this paper, we report the demonstration of a large-angle SC phenomenon based on pillar-type PCs. The light waves are accepted into these pillar-type PCs from a wide-range of incident angles and can be highly localized along the rod arrays. A propagation length of 0.4 mm was observed over a wavelength range of 1540 nm to 1570 nm, which may be a sufficient propagation distance for on-chip photonic applications.

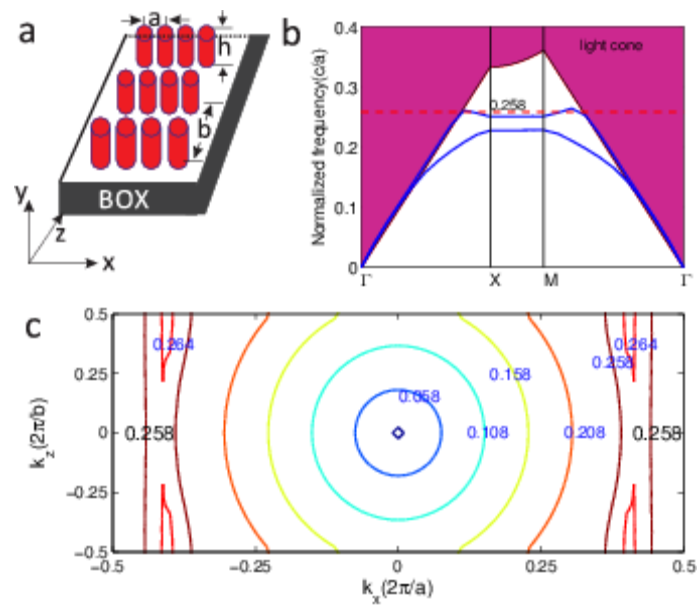

Fig. 1 a) Schematic of a 2D pillar-type PC consisting of a rectangular lattice formed by patterning a SOI wafer.

(b) The band diagram of a rectangular lattice PC. The dashed line represents the frequency $f = 0.258c/a$. The

colored region represents the light cone. (c) The equi-frequency contours for the second band.

## 2. Model and analysis

To increase the angular collimating range, one easy and feasible method is to lower the symmetry of the photonic structure by using a rectangular lattice [12]. Here, we consider 2D pillar-type PCs consisting of rectangular lattices fabricated by patterning a silicon-on-insulator (SOI) wafer, as shown in Fig. 1(a). A flat EFC across the entire Brillouin zone can be obtained by increasing the ratio of the side lengths of the rectangular lattice. However, an excessively large ratio will cause the light waves to leak to another direction (z direction) due to the small filling factor of the dielectric material. As a tradeoff between the large-angle self-collimation effect and the collimating efficiency, we selected the PC structure with the ratio b/a = 2.4. The top silicon layer with a height $h = 2.025a$ serves as the slab layer. The buried oxide (BOX) layer, together with air as the cladding layer, forms the high index-contrast structure required to achieve out-of-plane light confinement due to total internal reflection. The refractive indices of the silicon slab and BOX are $n_{Si}$ = 3.5 and $n_{SiO2}$ = 1.5, respectively. The band diagrams for the TM-like polarization (with the E-field along the y direction) are obtained using the plane wave expansion (PWE) method, as shown in Fig. 1(b). The colored region is the light cone, below which are guided modes localized to the slab [10].

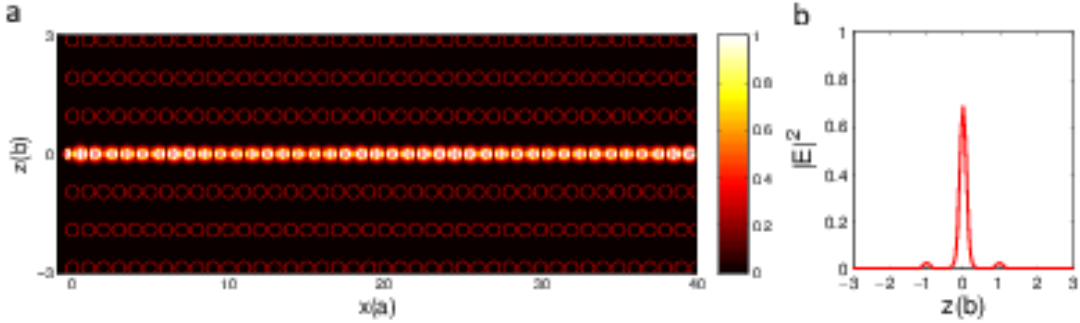

Fig. 2 (a) The field intensity of the electric field component $E_y$ (TM polarization) calculated using the FDTD method. (b) The $E_y$ field intensity along the z direction at x = 40a.

In PCs, the direction of light propagation completely depends on the gradient direction of the corresponding EFC because the direction of light propagation is identical with the group velocity $v_g = \nabla_k \omega(k)$, where $\omega$ is the optical frequency at the wave vector $k$ [19]. Therefore, the self-collimation effect can be considered to be due to the flat part of the EFC. However, this non-diffractive propagation is limited to functions using a rather wide beam, with the wave vectors restricted to the local flat portion of the EFC. For a narrow beam composed of a large range of $k$ vectors, i.e., to obtain all-angle self-collimation, a flat EFC across the entire first Brillouin zone is required. In our case, the corresponding EFC for the second band is calculated, as shown in Fig. 1(c). Note that the flat EFC at $f = 0.258c/a$ is flat across the first Brillouin zone, which is expected to result in large-angle self-collimation. To verify our prediction, a simulation is performed using the 3D finite-difference time-domain

(FDTD) method. A Gaussian beam with a width of $0.25b$ at $f = 0.258c/a$ is launched into the designed PC. The computed electric field intensity is plotted in Fig. 2(a). Self-collimation is achieved even though the light source contains a wide range of $k$ vectors due to the small beam circumference. In Fig. 2(b), we present the $E_y$ field intensity along the z direction at x = 40a. Note that the EM energy is highly localized along a single nanorod array. The $E_y$ field intensity in the third row to each side of this single nanorod array is only $1/10^5$, i.e., a very weak electric field. Placing four rows of nanorods to each side of the central single nanorod array will guarantee good confinement for the light waves.

## 3. Experimental results

For operation at optical communication frequencies, the parameters of the dielectric nanorod chain are designed to be $a = 400$ nm and $r = 160$ nm, and the height of the nanorod is $h = 810$ nm. Four main processing steps were used to fabricate the PCs on the platform of a silicon-on-insulator (SOI) wafer. First, thermal oxidation is required to form an oxide layer, which is used as a hard mask to transfer the pattern into the Si layer by reactive-ion etching (RIE). Second, electron-beam (E-beam) lithography is used to transfer the pattern to the high-resolution and high-sensitivity E-beam Resist ZEP-520. Third, we use RIE in a $CHF_3$ plasma to etch the hardmask. Finally, by using the thermally grown $SiO_2$ layer as a hard mask, the resulting dielectric rods are approximately 810 nm after using the inductively coupled plasma (ICP) etching process. As shown in the scanning electron micrograph (SEM) image of the nanorods in Fig. 3, the rods have a sidewall profile that is approximately 90° to the substrate; the screw thread is due to the Bosch-process of the deep etching.

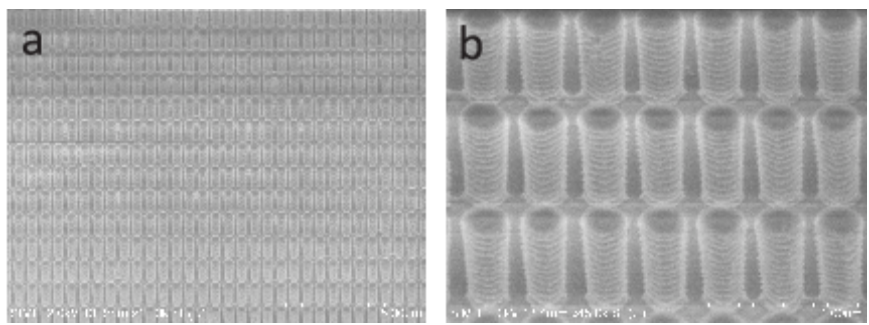

Fig. 3. SEM images of the fabricated device: (a) Rectangle lattice pillar-type PCs with a = 400 nm, b = 2.4a, h = 2.025a. (b) Magnified image.

In our experiment, to demonstrate the large-angle self-collimation effect, we designed three different samples with tilted PC regions. The light waves are introduced to illuminate the PCs by a waveguide with a width of 3 μm. The tilt angles relative to the incident waveguide are designed to be 0°, 45°, and 75°, as shown in the left panels (a), (b), and (c), respectively, of Fig. 4. Theoretically, self-collimation can be observed even for a nearly 90° incident angle. However, the larger incident angle results in more reflection loss due to the impedance mismatch. The maximum incident angle in our experiment is set to be 75° so that a sufficient amount of light scattering due to the imperfections of the structure can be captured with our camera. TM polarized light from a tunable laser source is introduced with a lensed fiber. To visualize the EM energy transport in the PCs, an infrared camera is used to capture the ray trace (via the scattered light) of the EM energy. Clear light ray traces are observed over the wavelength range of 1540 nm to 1570 nm. Coupled resonator waveguides (CROWs) can also provide a similar function to our PC structure of guiding light waves, but their operating frequencies are limited to the resonant frequencies of the whispering gallery modes of individual rods [22]. In contrast, our large-angle self-collimation effect is based on the dispersion relation of the whole PCs, which operates over a bandwidth that is broader than the resonant frequencies of individual rods. In Fig. 4 (d), (e), and (f), we present the captured results at a wavelength of 1550 nm that correspond to the incident angles of 0°, 45°, and 75°, respectively. The optical spot at point A is caused by the strong scattering at the interface of the waveguide facet and the PCs. The optical spot at point B is at the opposite end of the PC region, which also provides an interface for optical scattering. The route from A and B contains 1000 lattice periods, corresponding to a propagation length of 0.4 mm, which is a very considerable length scale for on-chip optical interconnection. The previous experimental results of the large-angle self-collimation effect indicate that light in the PC structures studied propagates at least 100 lattice periods without apparent divergence.[18] In our case, the propagation length can be up to 1000 lattice periods, with a more clearly indicated ray trace of the light beam than the previous results.

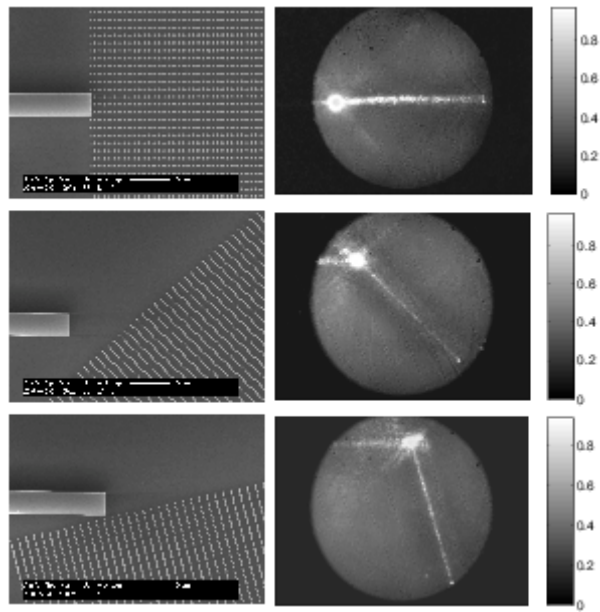

Fig. 4. Left panels: SEM pictures of designed PC samples with tilt angles of (a) 0°, (b) 45°, and (c) 75°. Right panels (d), (e), and (f): top views of the scattered light indicating the ray trace of the light beam captured by an infrared camera. The optical spots due to strong scattering at the interface are marked with A and B.

For quantitative results, the field intensity along the nanorod array determined from the FDTD calculations [23] is plotted in Fig. 5(a), and the theoretical propagation loss is determined to be approximately 10.7 dB/mm. The experimentally determined propagation loss of the self-collimated beam is measured by plotting the relative intensity along the nanorods in the image in Fig. 4(d). Fig. 5 shows that the experimental propagation loss is determined to be 17.6 dB/mm, which is higher than the simulation result. The additional propagation loss is believed to be primarily due to the roughness of the pillars, which can be reduced by improving the fabrication process. According to our simulation, the coupling efficiency of the light waves from air to the rod structure can be up to 92%, 90%, and 45% for incident angles of 0°, 45°, and 75°, respectively. In addition, the coupling efficiency can be enhanced by depressing the scattering at point A through the modification of the interface [20], [21]. Therefore, we believe that with the improvements mentioned above, a millimeter-scale propagation length can be easily observed with the rod structure.

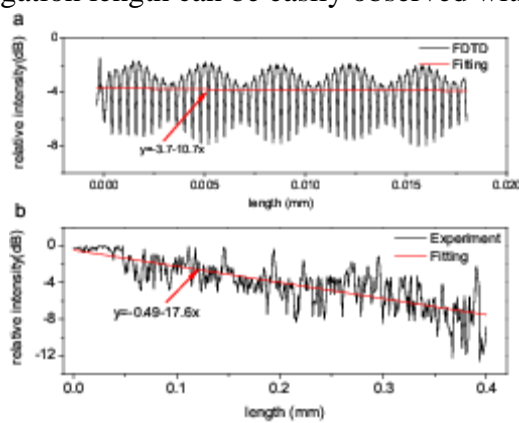

Fig. 5. (a) Theoretical propagation loss from the 3D FDTD simulation. (b) Experimental propagation loss of the self-collimated beam determined by plotting the relative intensity (in logarithmic scale) along the nanorods in Fig. 4(d).

## 4. Conclusion

To conclude, we designed, fabricated and characterized a structure based on silicon nanorods that exhibits a large-angle self-collimation phenomenon. The millimeter-scale propagation length and broad wavelength range exhibited by our structure may be sufficient for on-chip photonic applications. Such a large-angle self-collimation phenomenon resulting from the nanorod structure may also find applications in sensing and optofluidics.

## Acknowledgement

This work was supported in part by the National Natural Science Foundation of China under Grants 61107031, 11104305, 11204340, and 61275112; by the 863 Project under Grant 2012AA012202; and by the Science and Technology Commission of Shanghai Municipality underGrants 10DJ1400400, 11ZR1443700, and 11ZR1443800.